\documentclass[aps,prl,reprint,10pt,a4paper,preprintnumbers]{revtex4-1}
\usepackage{amsmath,amssymb,latexsym}
\usepackage{bm}
\usepackage{braket}
\usepackage{graphicx}
\usepackage{hepunits}
\usepackage{color}

\begin{document}

\title{Resummed differential cross sections for top-quark pairs at the LHC}

\author{Benjamin D. Pecjak$^b$, Darren J. Scott$^b$, Xing Wang$^a$, Li Lin Yang$^{a,c,d}$}

\affiliation{$^a$School of Physics and State Key Laboratory of Nuclear Physics and Technology, Peking University, Beijing 100871, China \\
$^b$Institute for Particle Physics Phenomenology, University of Durham, DH1 3LE Durham, UK
\\
$^c$Collaborative Innovation Center of Quantum Matter, Beijing, China
\\
$^d$Center for High Energy Physics, Peking University, Beijing 100871, China}

\begin{abstract}
We present state of the art resummation predictions for differential cross sections in top-quark pair production at the LHC. They are derived from a formalism which allows the simultaneous resummation of both soft and small-mass logarithms, which endanger the convergence of fixed-order perturbative series in the boosted regime, where the 
partonic center-of-mass energy is much larger than the mass to the top quark. We combine such a double resummation at NNLL$'$ accuracy with standard soft-gluon resummation at NNLL accuracy and with NLO calculations, so that our results are applicable throughout the whole phase space. We find that the resummation effects on the differential distributions are significant, bringing theoretical predictions into better agreement with experimental data compared to fixed-order calculations.  Moreover, such effects are not well described by the NNLO approximation
of the resummation formula, especially in the high-energy tails of the distributions, highlighting the importance of 
all-orders resummation in dedicated studies of boosted top production. 
\end{abstract}

\preprint{IPPP/16/04}

\maketitle

\section{Introduction}

The \unit{8}{\TeV} run of the LHC delivered about \unit{20}{\invfb} of
integrated luminosity to both the ATLAS and CMS experiments. Among
the many important results coming from these data, the properties of
the top-quark have been measured with unprecedented precision. At the
same time, theoretical calculations of top-quark related observables
have seen significant advancements in the last few years. In
particular, very recently the next-to-next-to-leading order (NNLO) QCD
corrections to differential cross sections in top-quark pair
($t\bar{t}$) production have been calculated \cite{Czakon:2015owf}. In
\cite{CMS-PAS-TOP-15-011}, the CMS collaboration performed a
comprehensive comparison between their measurements
\cite{Khachatryan:2015oqa} of the differential cross sections and
various theoretical predictions, including those from the NNLO
calculation and those from Monte Carlo event generators with
next-to-leading order (NLO) accuracy matched to parton showers. The
overall agreement between theory and data is truly remarkable, which
adds to the success of the Standard Model (SM) as an effective
description of Nature.

However, a persistent issue in the 8~TeV results is that the
transverse momentum ($p_T$) distribution of the top quark is softer in
the data than in theoretical predictions, i.e., the experimentally
measured differential cross section at high $p_T$ is lower than
predictions from event generators or from NLO fixed-order calculations
\cite{Khachatryan:2015oqa, Aad:2015hna}. While the NNLO corrections
bring the fixed-order predictions into better agreement with the CMS data, as
noted in \cite{Czakon:2015owf} and \cite{CMS-PAS-TOP-15-011}, there is
still some discrepancy in the high-$p_T$ bins where $p_T >
\unit{200}{\GeV}$.  Given the importance of the $t\bar{t}$ production
process as a standard candle for validating the SM and as an essential
background for new physics searches, it would be disconcerting if this
feature were to persist at higher $p_T$ values and with more data. It
is therefore important to assess the effects of QCD corrections even
beyond NNLO, in order to see whether the gap between theory and data
at high $p_T$ can be bridged.

For boosted top-quark pairs with high $p_T$ there are two classes of
potentially large contributions. The first is the Sudakov-type double
logarithms arising from soft gluon emissions. The second comes from
gluons emitted nearly parallel to the top quarks, resulting in large
logarithms of the form $\ln^n(m_t/m_T)$, where $m_t$ is the top quark
mass, and $m_T \equiv \sqrt{m_t^2+p_T^2}$ is the transverse mass of
the top quark. In \cite{Ferroglia:2012ku}, some of the authors of the
current work developed a formalism for the simultaneous resummation of
both type of logarithms to all orders in the strong coupling constant
$\alpha_s$. In this Letter, we report the first phenomenological
applications of that formalism, giving predictions for the top-quark
$p_T$ and the $t\bar{t}$ invariant mass distributions at the
\unit{8}{\TeV} LHC, and comparing with experimental measurements as
well as the NNLO calculations when possible. With an eye to the
future, we also present predictions for the \unit{13}{\TeV} LHC, where
NNLO results are not yet available.

Our main finding is that the higher-order effects contained in our
resummation formalism significantly alter the high-energy tails of the
$p_T$ and $t\bar{t}$ invariant mass distributions, softening that of the
$p_T$  distribution but enhancing that of the $t\bar{t}$ invariant mass
distribution. These effects bring our results into better 
agreement with the experimental data compared to pure NLO
fixed-order calculations. Interestingly, for the case of the $p_T$ distribution, this
softening of the spectrum is slightly stronger than the
similar effect displayed in recent NNLO results, and leads to
a better modeling of the $p_T>200$~GeV portion of the CMS data
\cite{Khachatryan:2015oqa}. We comment further on this fact in the 
conclusions.

\section{Formalism}

Our predictions are based on the factorization and resummation formula
derived in \cite{Ferroglia:2012ku}. The technical details will be given in a
forthcoming article, although the main elements have already been sketched out
in \cite{Ferroglia:2015ivv}.  In the kinematic situation where the top
quarks are highly boosted and the events are dominated by soft gluon
emissions, the resummed partonic differential cross section in Mellin
space can be written as
\begin{align}
\label{eq:boosted-resummed}
\widetilde{c}_{ij}(N,M_{t\bar{t}},m_t,\mu_f) &= \mathrm{Tr} \Bigg[ \widetilde{\bm{U}}_{ij}(\mu_f,\mu_h,\mu_s) \, \bm{H}_{ij}(M_{t\bar{t}},\mu_h) \nonumber
\\
&\hspace{-8em} \times \widetilde{\bm{U}}_{ij}^{\dagger}(\mu_f,\mu_h,\mu_s) \times \widetilde{\bm{s}}_{ij} \left( \ln\frac{M_{t\bar{t}}^2}{\bar{N}^2 \mu_s^2}, \mu_s \right) \Bigg]
\\
&\hspace{-8em} \times \widetilde{U}_D^2(\mu_f,\mu_{dh},\mu_{ds}) \, C_D^2(m_t,\mu_{dh}) \, \widetilde{s}_D^2 \Biggl( \ln\frac{m_t}{\bar{N}\mu_{ds}}, \mu_{ds} \Biggr) \, , \nonumber
\end{align}
where for simplicity, we have suppressed some variables in the functional
arguments which are unnecessary for the explanations below. In the above formula, $M_{t\bar{t}}$ is the invariant mass
of the $t\bar{t}$ pair (which can be related to the $p_T$ of the top quark in the soft 
limit through a change of variables), $N$ is the Mellin moment variable, and $\bar{N} \equiv
Ne^{\gamma_E}$ with $\gamma_E$ the Euler constant. The soft limit
corresponds to $N \to \infty$ in Mellin space. The four coefficient
functions $\bm{H}_{ij}$, $\widetilde{\bm{s}}_{ij}$, $C_D$ and
$\widetilde{s}_D$ encode contributions from four widely-separated
energy scales $M_{t\bar{t}}$, $M_{t\bar{t}}/\bar{N}$, $m_t$ and
$m_t/\bar{N}$, respectively. The presence of the four scales leads to
the two types of large logarithms discussed in the introduction. In correspondence
with these four physical scales, there are four unphysical
renormalization scales $\mu_h$, $\mu_s$, $\mu_{dh}$ and $\mu_{ds}$,
one for each coefficient function. The philosophy of resummation is to
choose the four unphysical scales to be around their corresponding
physical scales, so that the four coefficient functions are free of
large logarithms and are well-behaved in fixed-order perturbation
theory. One can then use renormalization group (RG) equations to
evolve these functions to the factorization scale $\mu_f$ in order to
convolute with the parton distribution functions (PDFs) and obtain the
hadronic cross sections. The effects of the RG running are encoded in
the two evolution factors $\widetilde{\bm{U}}_{ij}$ (for $\bm{H}_{ij}$
and $\widetilde{\bm{s}}_{ij}$) and $\widetilde{U}_D$ (for $C_D$ and
$\widetilde{s}_D$), which resum all the large logarithms to all orders
in $\alpha_s$ in an exponential form.

At the moment, the four coefficient functions are known to NNLO
\cite{Ferroglia:2012ku, Broggio:2014hoa, Ferroglia:2012uy}, while the
two evolution factors are known to next-to-next-to-leading logarithmic
(NNLL) accuracy \cite{Ferroglia:2012ku}. Such a level of accuracy is
usually referred to as NNLL$'$ in the literature, and we adopt that
nomenclature here. While the formula (\ref{eq:boosted-resummed}) is
only applicable in the boosted soft limit, we can extend its domain of
validity by combining it with information from NNLL soft gluon
resummation derived in \cite{Ahrens:2010zv} (recast into Mellin space)
as well as the NLO fixed-order result calculated in \cite{NLO} and
implemented in MCFM \cite{Campbell:2010ff}. The precise matching
formula can be found in \cite{Ferroglia:2015ivv}.  After such a
matching procedure, we denote the final accuracy of our predictions,
which are valid throughout phase space, as NLO+NNLL$'$. 

It would be desirable to match with the recent NNLO results in
\cite{Czakon:2015owf} to achieve NNLO+NNLL$'$ accuracy.  However, at
the moment NNLO results are only available for fixed (i.e.,
kinematics-independent) factorization and renormalization scales
$\mu_f \sim \mu_r \sim m_t$, whereas for the study of differential
distributions over large ranges of phase space we consider it
important to follow common practice and use dynamical (i.e.,
kinematics-dependent) scale choices. Therefore, such an improvement
over our result is not currently possible, and we leave it for the
future.

\section{Phenomenology}

In the following we present NLO+NNLL$'$ predictions for the
$M_{t\bar{t}}$ and $p_T$ distributions at the LHC. In all our numerics
we choose $m_t = \unit{173.2}{\GeV}$ and use MSTW2008NNLO PDFs
\cite{Martin:2009iq}. For $p_T$ distributions, the default values for
the factorization scale and the four renormalization scales are chosen
as $\mu_f = m_T$, $\mu_h = M_{t\bar{t}}$, $\mu_s =
M_{t\bar{t}}/\bar{N}$, $\mu_{dh} = m_t$ and $\mu_{ds} =
m_t/\bar{N}$. For $M_{t\bar{t}}$ distributions, the only difference is
$\mu_f = M_{t\bar{t}}$. We estimate scale uncertainties by varying the
five scales around their default values by factors of two and
combining the resulting variations of differential cross sections in
quadrature; we do not consider uncertainties from PDFs and $\alpha_s$
in this Letter. The hadronic differential cross sections are first
evaluated in Mellin space at a given point in phase space, and we then
perform the inverse Mellin transform numerically using the Minimal
Prescription \cite{Catani:1996yz}. This procedure relies on an
efficient construction of Mellin-transformed parton luminosities, for
which we use methods outlined in \cite{Bonvini:2014joa,
  Bonvini:2012sh}.

The differential cross sections considered below span several orders
of magnitude when going from low to high values of $p_T$ or
$M_{t\bar{t}}$. In order to better display the relative sizes of
various results, we show in the lower panel of each plot the
differential cross sections normalized to our default prediction,
i.e., the ratio defined by
\begin{align}
\text{Ratio} \equiv \frac{d\sigma}{d\sigma^{\text{NLO+NNLL$'$}}\big(\mu_i = \mu_i^{\text{default}}\big)} \, .
\end{align}

\begin{figure}[t!]
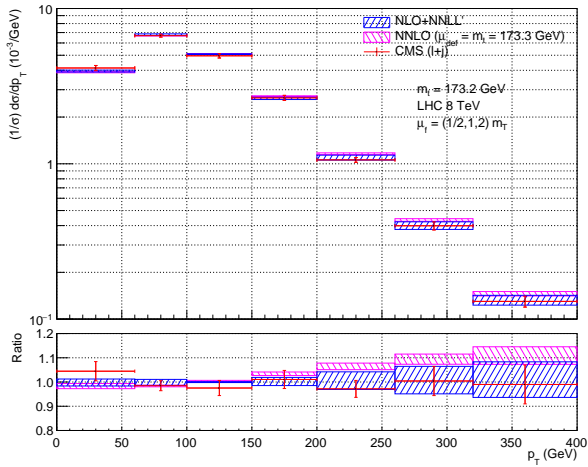

	\centering
	\includegraphics[width=0.9\linewidth]{{{pTt_lhc8_cms_0.50}}}
	\vspace{-3ex}
	\caption{\label{fig:pTt_cms}Resummed prediction (blue band) for the normalized top-quark $p_T$ distribution at the 8~TeV LHC compared with CMS data (red crosses) \cite{Khachatryan:2015oqa} and the NNLO result (magenta band) \cite{Czakon:2015owf}. 
The lower panel shows results normalized to the default NLO+NNLL$'$ prediction.}
\end{figure}

Fig.~\ref{fig:pTt_cms} compares our NLO+NNLL$'$ resummed prediction
for the normalized top-quark $p_T$ distribution to the CMS measurement
\cite{Khachatryan:2015oqa} in the lepton+jet channel at the LHC with a
center-of-mass energy $\sqrt{s}=\unit{8}{\TeV}$. Also shown is the
NNLO result from \cite{Czakon:2015owf}, which adopted by default the
renormalization and factorization scales $\mu_r=\mu_f=m_t$, and also
used a slightly different top-quark mass, $m_t=\unit{173.3}{\GeV}$. At
low $p_T$, it is clear that both the NLO+NNLL$'$ and the NNLO results
describe the data fairly well. With the increase of $p_T$, it appears
that the NNLO prediction systematically overestimates the data,
although there is still agreement within errors.
On the other hand, with the simultaneous resummation of the soft gluon
logarithms and the mass logarithms and also with the dynamical scale
choices, our NLO+NNLL$'$ resummed formula produces a softer spectrum
which agrees well with the data.

\begin{figure}[t!]
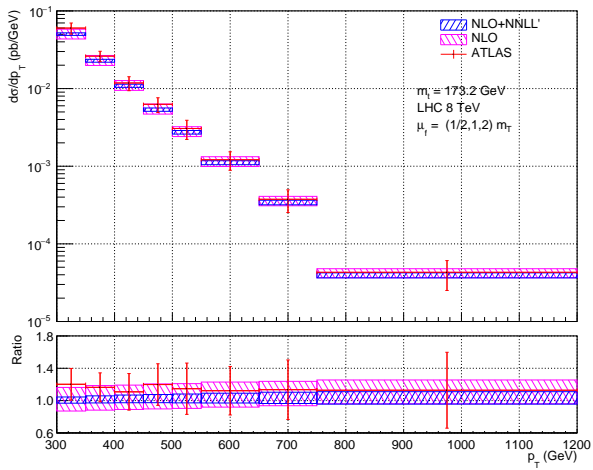

	\centering
	\includegraphics[width=0.9\linewidth]{{{pTt_lhc8_atlas_boosted_0.50}}}
	\vspace{-3ex}
	\caption{\label{fig:pTt_atlas_boosted}Resummed prediction (blue band) for the absolute $p_T$ distribution at the \unit{8}{\TeV} LHC in the boosted region compared with the ATLAS data (red crosses) \cite{Aad:2015hna} and the NLO result (magenta band).}
\end{figure}

\begin{figure}[t!]
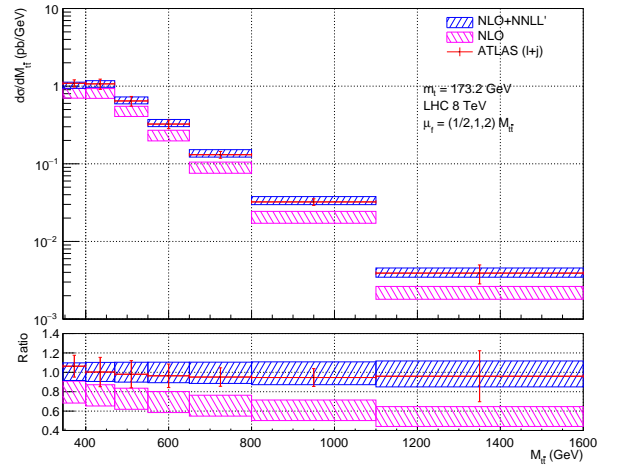

\centering
\includegraphics[width=0.9\linewidth]{{{Mtt_lhc8_atlas_1.00}}}
\vspace{-3ex}
\caption{\label{fig:Mtt_atlas}Resummed prediction (blue band) for the absolute $M_{t\bar{t}}$ distribution at the \unit{8}{\TeV} LHC compared with ATLAS data (red crosses) \cite{Aad:2015mbv} and the NLO result (magenta band).}
\end{figure}

In \cite{Aad:2015hna}, the ATLAS collaboration carried out a
measurement of the top-quark $p_T$ spectrum in the highly-boosted
region using fat-jet techniques. Although the experimental uncertainty
is rather large due to limited statistics, it is interesting to
compare it with the theoretical predictions here, since it is expected
that the soft and small-mass logarithms become more relevant at higher
energies. In Fig.~\ref{fig:pTt_atlas_boosted} we show such a
comparison. The NNLO result for such high $p_T$ values is not yet
available, so we compare instead with the NLO result computed using
MCFM with MSTW2008NLO PDFs and dynamical renormalization and
factorization scales, whose default values are
$\mu_r=\mu_f= m_T$.  Scale uncertainties of the NLO results
are estimated through variations of $\mu_r=\mu_f$ by a factor of two
around the default value.  From the plot one can see that the NLO
result calculated in this way does a good job in estimating the
residual uncertainty from higher order corrections, as the resummed
band lies almost inside the NLO one up to $p_T = \unit{1.2}{\TeV}$. On
the other hand, the inclusion of the higher-order logarithms in the
NLO+NNLL$'$ result significantly reduces the theoretical uncertainty,
which is crucial for future high precision experiments at the LHC.

\begin{figure}[t!]
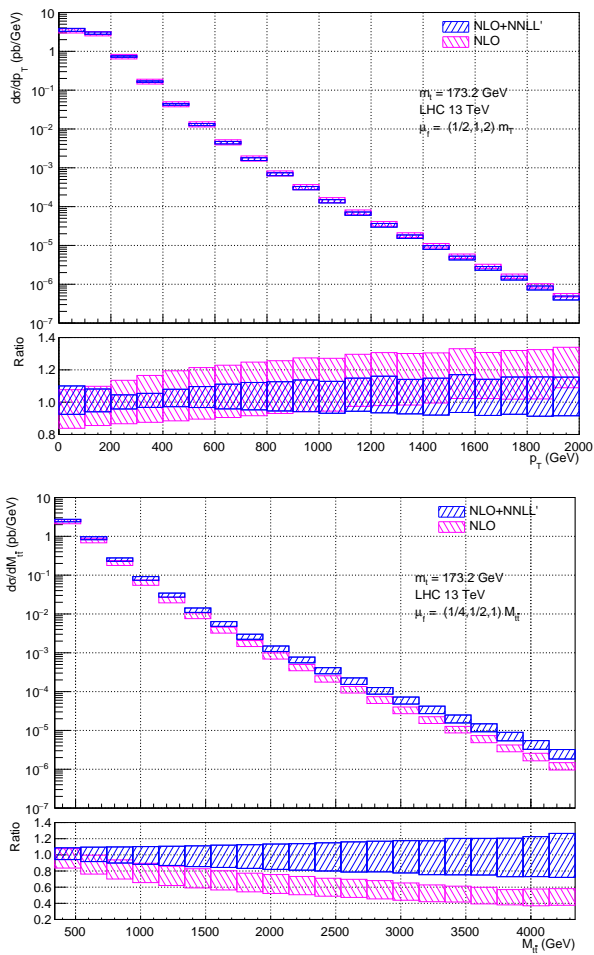

	\centering
	\includegraphics[width=0.9\linewidth]{{{pTt_lhc13_0.50}}}
	\\
	\includegraphics[width=0.9\linewidth]{{{Mtt_lhc13_0.50}}}
	\vspace{-3ex}
	\caption{\label{fig:lhc13}Resummed predictions (blue bands) for the $p_T$ and $M_{t\bar{t}}$ distributions at the \unit{13}{\TeV} LHC compared with the NLO results (magenta bands).}
\end{figure}

Our formalism is flexible and can be applied to other differential
distributions as well. To demonstrate this fact, in
Fig.~\ref{fig:Mtt_atlas} we show the NLO+NNLL$'$ resummed prediction
for the top-quark pair invariant mass distribution along with a
measurement from the ATLAS collaboration \cite{Aad:2015mbv} at the
\unit{8}{\TeV} LHC. Since the NNLO result in
\cite{Czakon:2015owf} for this distribution has an incompatible binning, 
it is currently not possible to include it in the plot, so we show
instead the NLO result computed with the same input as in Fig.~\ref{fig:pTt_atlas_boosted}, but this
time with the default scale choice $\mu_r=\mu_f=M_{t\bar{t}}$.  One
can see from the plot that the NLO result with this scale choice is
consistently lower than the experimental data.  The resummation
effects significantly enhance the differential cross sections,
especially at high $M_{t\bar{t}}$. As a result, the NLO+NNLL$'$
prediction agrees with data quite well.  We have found that 
choosing the default renormalization and factorization scales
to be half the invariant mass increases the fixed-order cross section and therefore mimics
to some extent the resummation effects.  In fact, this procedure has been extensively employed in the literature 
for processes such as Higgs production \cite{Anastasiou:2015ema}, where 
higher-order corrections are also large.  Consequently, it may be advisable to
employ a renormalization and factorization scale of the order of
$M_{t\bar{t}}/2$ in fixed-order calculations (and Monte Carlo event
generators), and  we shall use this choice when studying the $M_{t\bar t}$
distribution at the 13~TeV LHC below. 

\begin{figure}[t!]
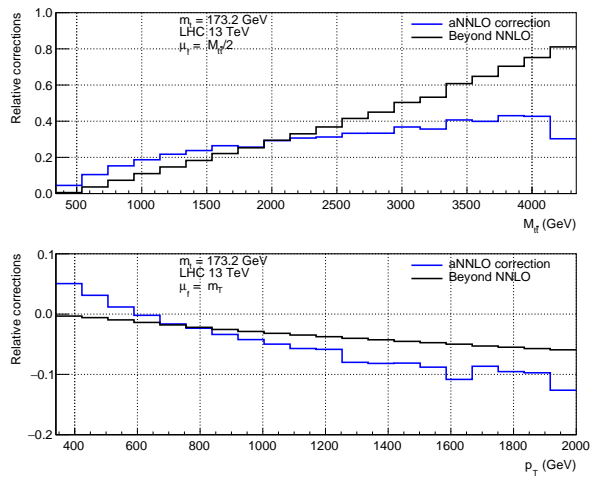

	\centering
	\includegraphics[width=0.9\linewidth]{{{beyond_nnlo_lhc13}}}
  \vspace{-3ex}
  \caption{\label{fig:lhc13-radcors} Relative sizes of the corrections
    at approximate NNLO (blue) and beyond (black), with respect to NLO. See
    Eq.~(\ref{rat-defs}) and the explanations there for precise
    definitions.}
\end{figure}

The LHC has started the \unit{13}{\TeV} run in 2015. So far there are
only two CMS measurements \cite{CMS:2015toa, CMS:2015bta} of
differential cross sections for $t\bar{t}$ production, based on just
\unit{42}{\invpb} of data. The resulting experimental uncertainties
are therefore quite large and it is not yet possible to probe higher
$p_T$ or $M_{t\bar{t}}$ values. Nevertheless, in the near future there
will be a large amount of high-energy data, which will enable
high-precision measurements of $t\bar{t}$ kinematic distributions,
also in the boosted regime. In Fig.~\ref{fig:lhc13} we show our
predictions for the $p_T$ and $M_{t\bar{t}}$ spectrum up to $p_T =
\unit{2}{\TeV}$ and $M_{t\bar{t}} = \unit{4.34}{\TeV}$, contrasted
with the NLO results. Note that for the $M_{t\bar{t}}$ distribution,
we have changed the default $\mu_f$ to a lower value $M_{t\bar{t}}/2$
for the reasons explained above. The plots exhibit similar patterns as
observed at $\unit{8}{\TeV}$, namely that the higher-order resummation
effects serve to soften the tail of the $p_T$ distribution but enhance
that of the $M_{t\bar{t}}$ distribution compared to a pure NLO
calculation.

As mentioned before, we would like to match our calculations with the 
NNLO results when they become available in the future. 
We end this section by discussing the expected effects of such a matching, 
by estimating the size of resummation corrections beyond NNLO.
We do this in Fig.~\ref{fig:lhc13-radcors}, where the relative sizes
of the beyond-NNLO corrections generated through the resummation
formula are displayed as a function of $M_{t\bar{t}}$ or $p_T$ with
the default scale choices. The exact NNLO results for these scale
choices are not yet available, so we show in comparison the relative
sizes of the approximate NNLO (aNNLO) corrections obtained by expanding and
truncating our resummation formula to that order. More precisely, the
blue and black curves in Fig.~\ref{fig:lhc13-radcors} correspond to
\begin{align}
\label{rat-defs}
\text{aNNLO correction} &\equiv 
\frac{d\sigma^{\text{aNNLO}}-d\sigma^{\text{NLO}}}{d\sigma^{\text{NLO}}} \, 
, 
\\
\text{Beyond NNLO}& \equiv 
\frac{d\sigma^{\text{NLO+NNLL$'$}}- d\sigma^{\text{aNNLO}}}{d\sigma^{\text{NLO}}} 
\nonumber \, ,
\end{align}
where $d\sigma^{\text{aNNLO}}$ refers to the approximate NNLO result. The figure
clearly shows that corrections beyond NNLO are significant in the
tails of the distributions, especially in the case of the $M_{t\bar
  t}$ distribution.

\section{Conclusions and outlook}

In this Letter we have presented new resummation predictions for differential cross sections in $t\bar{t}$ production at the LHC. The predictions include the simultaneous resummation to NNLL$'$ accuracy of both soft and small-mass logarithms, which endanger the convergence of
the fixed-order perturbative series in the boosted regime where the partonic center-of-mass energy is much larger than the mass of the top quark. This resummation is matched with both standard soft-gluon resummation at NNLL accuracy and  fixed-order NLO calculations, so that our results are applicable in the whole phase space.  
Such predictions for $t\bar{t}$ differential distributions at the LHC are not only the first to be calculated in Mellin space, but also represent the highest resummation accuracy achieved to date, namely NLO+NNLL$'$. 
Our results are thus a major step forward in the modeling of high-energy tails of distributions, which is 
of great importance for new physics searches.  

The agreement of NLO+NNLL$'$ predictions with data  indicates the value of including resummation effects and using dynamical scale settings correlated with $p_T$ or $M_{t\bar{t}}$ when studying differential distributions.
Interestingly, in the case of normalized $p_T$ distribution measured by the CMS collaboration~\cite{Khachatryan:2015oqa}, the NLO+NNLL$'$ calculation produces a slightly softer spectrum than recent NNLO predictions (which use a fixed scale setting where $\mu_f=\mu_r=m_t$ by default), thus achieving a better agreement with the data. However, we emphasize that the optimal use of resummation is to supplement NNLO calculations, not to replace them. With this in mind, we have studied the size of corrections beyond NNLO encoded in our resummation formula, and found that their effects are significant in the high-energy tails of distributions, especially for the $t\bar{t}$ invariant mass distribution where they enhance the differential cross section.  It will therefore be an essential and informative exercise  
to produce NNLO+NNLL$'$ predictions once NNLO calculations are available with dynamical scale settings.

{\em Acknowledgments:\/}
We would like to thank Alexander Mitov for providing us the results of the NNLO calculations in \cite{Czakon:2015owf}.
We are grateful to Andrea Ferroglia for collaboration on many related works.
X.~Wang and L.~L.~Yang are supported in part by the National Natural Science Foundation of China under Grant No. 11575004.  D.~J.~Scott is supported by an STFC Postgraduate Studentship.

\end{document}